\documentclass[aps,pra,twocolumn,amsmath,amssymb,nofootinbib,showpacs,superscriptaddress]{revtex4-1}
\usepackage[english]{babel}
\usepackage{latexsym}
\usepackage{graphics}
\usepackage{graphicx}
\usepackage{epsfig}
\usepackage{color}
\usepackage{bm}
\usepackage{amsmath}
\usepackage{amssymb}
\usepackage{amsthm}
\usepackage{dcolumn}
\usepackage{bm}
\usepackage{float}
\usepackage{hyperref}
\usepackage{color}
\usepackage{epstopdf}
\usepackage{cleveref}
\usepackage[svgnames]{xcolor}
\hypersetup{hidelinks,colorlinks=true,allcolors=DarkBlue}

\begin{document}

\preprint{APS/123-QED}

\title{Symmetric blind information reconciliation and \\ hash-function-based verification for quantum key distribution}

\author{A.K. Fedorov}
\affiliation{Russian Quantum Center, Skolkovo, Moscow 143025, Russia}

\author{E.O. Kiktenko} 
\affiliation{Russian Quantum Center, Skolkovo, Moscow 143025, Russia}
\affiliation{Steklov Mathematical Institute of Russian Academy of Sciences, Moscow 119991, Russia}
\affiliation{Bauman Moscow State Technical University, Moscow 105005, Russia}

\author{A.S. Trushechkin}
\affiliation{Steklov Mathematical Institute of Russian Academy of Sciences, Moscow 119991, Russia}
\affiliation{National Research Nuclear University MEPhI, Moscow 115409, Russia}
\affiliation{Department of Mathematics and Russian Quantum Center, National University of Science and Technology MISiS, Moscow 119049, Russia}

\begin{abstract}
We consider an information reconciliation protocol for quantum key distribution (QKD).
In order to correct down the error rate, we suggest a method, which is based on symmetric blind information reconciliation for the low-density parity-check (LDPC) codes.
We develop a subsequent verification protocol with the use of $\epsilon$-universal hash functions, which allows verifying the identity between the keys with a certain probability.
\end{abstract}

\maketitle

\section{Introduction}

A paradigmatic problem of cryptography is the problem of key distribution~\cite{Schneier}.
Widely used tools for key distribution, based on so-called public key cryptography, 
use an assumption of the complexity of several mathematical problems such as integer factorization~\cite{RSA} and discrete logarithm~\cite{DH}.
However, a large-scale quantum computer would allow solving such tasks in a more efficient manner in compare with their classical counterparts using Shor's algorithm~\cite{Shor}. 
Therefore, most of information protection tools are vulnerable to attacks with the use of quantum algorithms. 
It should be also noted that absence of efficient non-quantum algorithms breaking public key tools remains unproved.

In the view of appearance of large-scale quantum computers, the crucial task is to develop a quantum-safe infrastructure for communications.  
Crucial components of such an infrastructure are information-theoretically secure schemes, which make no computational assumptions~\cite{Schneier}. 
Examples of these schemes are the one-time-pad encryption~\cite{Shannon,Vernam,Kotelnikov} and Wegman-Carter authentication~\cite{WegmanCarter1981}.
Nevertheless, the need for establishing shared secret symmetric keys between communicating parties invites the challenge of how to securely distribute keys. 

Quantum key distribution (QKD) offers an elegant method for key establishment between distant users (Alice and Bob), without relying on insecure public key algorithms~\cite{Gisin2002}.
During last decades, remarkable progress in theory, experimental study, and technology of QKD has been performed~\cite{Lo2016}. 
However, realistic error rates in the sifted key using current technologies are of the order of a few percent, which is high for direct applications~\cite{Gisin2002,Lo2016,Kiktenko2016}.
Practical QKD then include a post-processing procedure, which is based on the framework of the classical information theory.

In this work, we focus on an information reconciliation task in QKD.
We consider a method for error correction based on symmetric blind information reconciliation~\cite{Kiktenko2017} with low-density parity-check (LDPC) codes~\cite{LDPC,LDPC2}.
In order to check the identity between the keys after the error correction step, we develop a subsequent verification protocol with the use of $\epsilon$-universal hash functions~\cite{Krovetz}.
We note that the additional procedures (privacy amplification and authentication) 
are necessary to obtain secure keys shared only by Alice and Bob~\cite{Kiktenko2016,Kiktenko2017,KiktenkoAnufriev2016,KiktenkoTrushechkin2016}.

The present paper is organized as follows. 
In Sec.~\ref{sec:SBEC}, we describe the symmetric blind information reconciliation technique for the low-density parity-check (LDPC) codes.
In Sec.~\ref{sec:Verification}, we suggest a subsequent verification protocol with the use of $\epsilon$-universal hash functions, 
which allows one to verify the identity between the keys with a certain probability.
In Sec.~\ref{sec:Estimations}, we give estimations for the information leakage in the suggested information reconciliation protocol. 
We summarize the main results of our work in Sec.~\ref{sec:Conclusion}

\section{Symmetric blind error correction}\label{sec:SBEC}

\begin{figure}[t]
	\centering
	\includegraphics[width=1\linewidth]{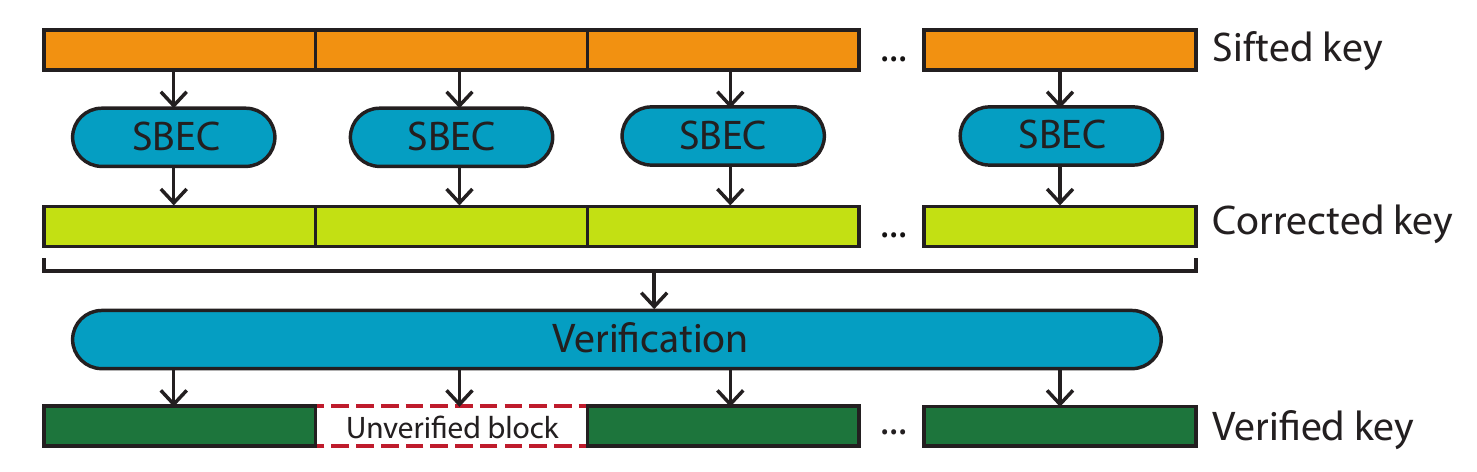}
	\vskip -2mm
	\caption{Scheme of the information reconciliation protocol:
	First, the block of the sifted key is split in sub-blocks which are treated with SBEC in parallel. 
	Second, all the sub-blocks go all together through the verification step. }
	\label{fig:ec1}
\end{figure}

We suggest the protocol for information reconciliation which has two essential steps: the symmetric blind error correction (SBEC) procedure and the subsequent verification protocol.
For the SBEC technique, a block of the sifted key is divided into $N_{\rm sb}$ sub-blocks of length $n_{\rm sb}$, and all sub-blocks are treated in parallel.
All the resulting sub-blocks from SBEC are input data for the subsequent verification procedure. 
If it is necessary during the verification protocol the keys are divided into the same sub-blocks once again for determining the sub-blocks which contain an error (for details, see Fig.~\ref{fig:ec1}).

Let us consider the SBEC procedure for the particular sub-blocks of the sifted key $k_{\rm sift}^{A}$ and $k_{\rm sift}^{B}$ of length $n_\mathrm{sb}$ owned by Alice and Bob.
For details of the SBEC procedure we refer the reader to Ref.~\cite{Kiktenko2017}, and here we confine ourself with explanation of the main steps.

For the implementation of the SBEC procedure we consider a set (``pool'') of nine LDPC codes~\cite{LDPC,LDPC2} with the following rates:
\begin{equation} \label{Rs}
	\mathcal{R}=\{0.9, 0.85, \ldots, 0.5\},
\end{equation} 
and the frame length $n_{\rm fr}$.
The parity-check matrices of these codes can be generated with the use of the improved progressive edge growing algorithm~\cite{MartnezMateo2010} 
according to generating polynomials from Ref.~\cite{Elkouss2009}.

The implementation of the SBEC procedure consists of following steps.

\begin{itemize}

	\item[0)]
	
This initial step of the SBEC procedure realizes the preliminary initialization. 
The parties initialize zero strings $e$ of length $n_{\rm fr}$.

	\item[1)]

Parties implement the rate adaptation.
Alice (Bob) extend their sub-blocks of the sifted key $k_{\rm sift}^{A(B)}$ of length
\begin{equation} \label{idnty}
	n_{\rm sb}:=0.95n_\mathrm{fr}
\end{equation}
with $n_{\rm shrt}$ shortened and $n_{\rm pnct}$ punctured symbols, where $n_\mathrm{shrt}:= \lceil h(q_\mathrm{est})n_\mathrm{fr} - n_\mathrm{sb}(1-R)\rceil$, $n_\mathrm{pnct}:=\Delta n_\mathrm{ext}-n_\mathrm{shrt}$.
Here, $\lceil\cdot\rceil$ stands for ceiling operation, 
$h(\cdot)$ is the binary entropy function, 
\begin{equation} \label{dnext}
	\Delta n_\mathrm{ext} := 0.05n_\mathrm{fr}
\end{equation}
is the total number of shortened and punctured symbols, 
and the code rate $R$ is chosen among $\mathcal{R}$ in such a way that both $n_\mathrm{shrt}$ and $n_\mathrm{pnct}$ are non-negative for current estimation of the QBER $q_{\rm est}$.

The positions for punctured symbols are chosen according untainted puncturing technique~\cite{Elkouss2012}, while the positions for shortened symbols are chosen pseudo-randomly.
We denote the list of positions with sifted key bits as $\Omega$.
	
	\item[2)]

This step is the realization of the syndromes exchange. 
Alice and Bob exchange with syndromes $s_{A(B)}:=k_{\rm ext}^{A(B)}\mathbf{H}_R^{\rm T}$,
where $\mathbf{H}_R$ is the parity-check matrix corresponded to code rate $R$ ($k_{\rm ext}^{A(B)}$ and $s_{A(B)}$ are treated as row-vectors; ${\rm T}$ stands for transposition). 
All the summations in vector matrix multiplication are assumed to be performed by modulo 2.

	\item[3)]
	
Alice and Bob use belief propagation decoding.
The parties apply syndrome decodings based on the belief propagation algorithm with log-likelihood ratio (LLR)~\cite{Kiktenko2017}.
As the input both the parties use the ``relative syndrome'' $\Delta s = s_A \oplus s_B$,
current information about error pattern $e$ (note, that it was initialized as zero string in the beginning if the procedure), 
parity-check matrix ${\bf H}_R$, 
the estimated QBER $q_{\rm est}$, 
and current positions of shortened and punctured symbols ($\oplus$ stands for modulo 2 summation).
If the algorithm converges, it returns the new value of the error pattern $e$, and then Alice calculates the corrected key as follows: $k_{\rm cor}^A:=k_{\rm ext}^A[\Omega]\oplus e[\Omega]$,
where $\Omega$ is the list of positions of sifted key symbols in the extended key.
Bob assumes his corrected key has the following form: $k_{\rm cor}^B:=k_{\rm sift}^B$.
The SBEC is finished.

	\item[4)]

If the belief propagation algorithm does not converge, the this step is applied.
It is based on disclosing additional information as follows. 
The parties take $d:=\lceil 56 - 40R \rceil$
positions in the extended key, which has the minimal magnitudes of LLR values to the end of belief propagation algorithm, 
disclose the values of their extended keys in these positions, and mark these positions as shortened.
Then the parties update their error patterns $e$ in the newborn shortened positions, and go to the Step 3.

\end{itemize}

There is still a certain probability that uncorrected errors remain after the SBEC procedure. 
In order to detect remaining errors, we implement the subsequent verification protocol, which is described below.

\section{Verification}\label{sec:Verification}

Here we suggest the verification protocol based on using the following $\epsilon$-universal family of hash functions~\cite{Krovetz}:
\begin{equation}\label{eq:verhash}
\begin{split}
	&h_k(X):=\text{\sffamily inttostr}\left[ \sum_{i=1}^{n}{\text{\sffamily strtoint}(x_i)k^{i-1}}~\mathrm{mod}~p \right],
\end{split}
\end{equation}
where $k\in \mathbb{F}\equiv\{0,1,2,\ldots,p-1\}$ is a randomly chosen key for the universal hashing, $p$ is the prime number, 
$X$ is a binary string of an arbitrary length, 
$(x_1 \| x_2 \| \ldots \| x_{n}):=X$ is a partition of the string into substrings $x_i$ of length $l_p = \lfloor \log_2p \rfloor$
($\lfloor\cdot\rfloor$ stands for floor operation), 
{\sffamily inttostr} and {\sffamily strtoint} are functions performing conversion between integer values an binary strings.

It can be shown~\cite{Krovetz} that the collision probability of the hash function~(\ref{eq:verhash}) (\emph{i.e.} the probability that $h_k(X)=h_k(Y)$ for some $X\neq Y$ and random $k$) is given by following expression: $\epsilon(l)\leq (\lceil l/l_p \rceil-1)/p$,
where $l$ is length of $X$.
Below the verification protocol based on the considered family of universal hash functions is considered.

Consider blocks of the corrected keys $K_{\rm cor}^{A(B)}$ owned by Alice (Bob), which consist of $n_{\rm b}=N_{\rm sb}\times n_{\rm sb}$ bits.
The implementation of the suggested verification protocol consists of following steps.

\begin{itemize}

	\item[1)]
On the initial step of the verification protocol, the parties use generation of the key for universal hashing.
Alice generate a random number $k\in\mathbb{{F}}$ using a true random generator (TRNG).
	
	\item[2)]
Further, the calculation of the hash for the whole block on the Alice side is realized. 
Alice computes $h_k(K_{\rm cor}^A)$ and sends it Bob together with $k$.
	
	\item[3)]	
This step is comparing the hashes for the whole block
Bob computes $h_k(K_{\rm cor}^B)$ and compares it with $h_k(K_{\rm cor}^A)$.
If the hashes are identical, then Bob sends the acknowledgement message to Alice.
The parties then assume $K_{\rm ver}^{A(B)}:=K_{\rm cor}^{A(B)}$, 
and the protocol finishes.
Here $K_{\rm sift}^{A(B)}$ are blocks owned by Alice (Bob).
If the hashes are different, then Bob sends the negative-acknowledgement message, and the protocol continues.

	\item[4)]	
Computing the hashes for all the sub-blocks on the Alice side is Step 4.
Alice splits the corrected key $K^A_{\rm cor}$ in $N_{\rm sb}$ sub-blocks $\{k^A_{{\rm cor},i}\}$ of length $n_{\rm sb}$, 
generates $N_{\rm sb}$ random keys $\{k_i\}$ belonging to $\mathbb{F}$, calculates $\{h_{k_i}(k^A_{{\rm cor},i})\}$, and sends both these sets to Bob.

	\item[5)]	
Comparing the hashes for all the sub-block is used as the final step. 
Bob computes his versions of hashes $\{h_{k_i}(k^B_{{\rm cor},i})\}$, 
compares them with corresponding values from Alice, 
and discard the sub-blocks with mismatched hashes from his corrected key $K_{\rm cor}^B$, to obtain the verified key $K_{\rm ver}^B$.
Then he sends Alice the indices of unverified block, and she perform the same operation and the protocol finishes.

\end{itemize}

To estimate a probability of the remaining error in the verified keys consider a worst case scenario where all of $N_\mathrm{sb}$ sub-blocks contain errors after SBEC procedure.
In this case, the probability of at least hash collision is given by the following expression:
\begin{equation}
\begin{split}
	\epsilon_\mathrm{ver}&\leq\epsilon(n_\mathrm{b})+\left[1-\epsilon(n_\mathrm{b})\right]\left[1-(1-\epsilon(n_\mathrm{sb}))^{N_\mathrm{sb}}\right].
\end{split}
\end{equation}
Here the first term is a probability of collision for the whole block, and the second term is a probability of at least one collision in the verification of all the sub-blocks in the case of different hashes of the whole blocks . 
We note that the initial processing of the whole blocks $K_{\rm cor}^{A}$ and $K_{\rm cor}^{B}$
is performed to minimize the leakage of the information via public discussion.
Let the frame error rate (FER), that is a probability of remaining error after SBEC in each of the sub-blocks, to be equal to $F$.
Then the information leakage in the verification step, neglecting the hash collision event, is given by the following expression: 
\begin{multline} \label{leak}
	{\rm leak}_{\rm ec}^{\rm ver} = (1-F)^{N_{\rm sb}}l_{\rm ht}+\\+\left[1-(1-F)^{N_{\rm sb}}\right](N_{\rm sb}+1)l_{\rm ht},
\end{multline} 
where
\begin{equation}
	l_{\rm ht}=\lceil \log_2p \rceil
\end{equation}
is a hash length.
In Eq.~\eqref{leak} the first term corresponds to the case, where the whole blocks are identical and only one verification hash is transfered.
The second term corresponds to the case of at least one error, where additional $N_{\rm sb}$ hashes are transfered.

\section{Estimations}\label{sec:Estimations}

Let us consider our protocol based on a set of LDPC codes of frame length $n_{\rm fr}=4000$ and code rates given by Eq.~\eqref{Rs}.
According to Eq.~\eqref{dnext} and Eq.~\eqref{idnty} the number of sifted key bit processed in each sub-block is $n_{\rm sb}=3800$.
Taking the number of sub-blocks $N_{\rm sb}=256$ and the prime number for universal hashing $p=2^{50}-27$ with hash length $l_{\rm ht}=50$ bit, 
we obtain the following bound on a probability of the verification fail: $\epsilon_{\rm ver}\leq 2\times 10^{-11}$.

Let us then consider a question about information leakage in the verification step.
Assuming that the FER of SBEC is $F=10^{-5}$, we obtain the following result: ${\rm leak}_{\rm ec}^{\rm ver} \approx 1.65 l_{\rm ht} \approx 83\, {\rm bit}$.

In order to compare our approach with currently available post-processing tools,
we calculate the information leakage for the setups described in Ref.~\cite{Gisin3}.
We note that in this case the hashes are added to each processed LDPC code block.
Therefore, one has the following estimation: ${\rm leak}_{\rm ec}^{\rm ver,alt} \approx n_{\rm} l_{\rm ht} =12\,800\,{\rm bit}$.

It is thus clearly seen that the suggested approach has an advantage in ${\rm leak}_{\rm ec}^{\rm ver,alt} / {\rm leak}_{\rm ec}^{\rm ver}\approx{155}\mbox{ times}$.
Thus, the suggested information reconciliation protocol allows one to decrease the information leakage in the verification protocol significantly.

\section{Conclusion}\label{sec:Conclusion}

We have presented the information reconciliation protocol which combines two approaches: SBEC, based on LDPC codes, and verification, based on $\epsilon$-universal hashing.
We have shown that applying SBEC for a number of sub-blocks in parallel and performing verification for the general block allows significant decreasing the information leakage in the verification stage.

The presented procedure allows one to obtain identical keys from sifted keys with a known bound of an error probability, which depends on the particular parameters of the protocol, 
such as hash length, block length, and number of sub-blocks.

The open source proof-of-principle realization of the presented algorithms are available~\cite{KiktenkoAnufriev2016,KiktenkoTrushechkin2016}.

{\bf Acknowledgments}. We thank Y.V.~Kurochkin and N.O.~Pozhar for useful discussions. 
The work of A.T. and E.K. was supported by the grant of the President of the Russian Federation (project MK-2815.2017.1).
A.K.F. is supported by the RFBR grant (17-08-00742).


\begin{thebibliography}{}

\bibitem{Schneier}
B. Schneier,
{\it Applied cryptography} 
(John Wiley \& Sons, Inc., New York, 1996).

\bibitem{RSA}
R.L. Rivest, A. Shamir, and L. Adleman, 
{\href{https://dx.doi.org/10.1145/359340.359342}{Commun. ACM {\bf 21}, 120 (1978)}}.

\bibitem{DH}
W. Diffie and M.E. Hellman,
{\href{https://dx.doi.org/10.1109/TIT.1976.1055638}{IEEE Trans. Inform. Theor. {\bf 22}, 644 (1976)}}.

\bibitem{Shor}
P.W. Shor,
{\href{https://dx.doi.org/10.1137/S0036144598347011}{SIAM J. Comput. {\bf 26}, 1484 (1997)}}.

\bibitem{Vernam}
G.S. Vernam,
{\href{https://www.doc.ic.ac.uk/~mrh/330tutor/vernam.pdf}{J. Amer. Inst. Electr. Engineers {\bf 45}, 109 (1926)}}.

\bibitem{Shannon}
C.E. Shannon,
{\href{https://dx.doi.org/10.1002/j.1538-7305.1948.tb01338.x}{Bell Syst. Tech. J. {\bf 27}, 379 (1948)}}.

\bibitem{Kotelnikov}
V.A. Kotel'nikov,
Classified Report (1941); 
see S.N. Molotkov, 
{\href{https://dx.doi.org/10.1070/PU2006v049n07ABEH006050}{Phys. Usp. {\bf 49}, 750 (2006)}}.

\bibitem{WegmanCarter1981}
M.N. Wegman and J.L. Carter,
{\href{https://dx.doi.org/10.1016/0022-0000(81)90033-7}{J. Comp. Syst. Sci. {\bf 22}, 265 (1981)}}.

\bibitem{Gisin2002}
N. Gisin, G. Ribordy, W. Tittel, and H. Zbinden,
{\href{https://dx.doi.org/10.1103/RevModPhys.74.145}{Rev. Mod. Phys. {\bf 74}, 145 (2002)}}.

\bibitem{Lo2016}
E. Diamanti, H.-K. Lo, and Z. Yuan, 
{\href{https://dx.doi.org/10.1038/npjqi.2016.25}{npj Quant. Inf. {\bf 2}, 16025 (2016)}}.

\bibitem{Kiktenko2016}
E.O. Kiktenko, A.S. Trushechkin, Y.V. Kurochkin, and A.K. Fedorov,
{\href{https://dx.doi.org/10.1088/1742-6596/741/1/012081}{J. Phys. Conf. Ser. {\bf 741}, 012081 (2016)}}.

\bibitem{Kiktenko2017}
E.O. Kiktenko, A.S. Trushechkin, C.C.W. Lim, Y.V. Kurochkin, and A.K. Fedorov,
{\href{https://dx.doi.org/10.1103/PhysRevApplied.8.044017}{Phys. Rev. Applied {\bf 8}, 044017 (2017)}}.

\bibitem{KiktenkoAnufriev2016}
E.O. Kiktenko, M.N. Anufriev, N.O. Pozhar, and A.K. Fedorov,
{\href{https://dx.doi.org/10.5281/zenodo.164953}{Symmetric information reconciliation for the QKD post-processing procedure (2016)}}.

\bibitem{KiktenkoTrushechkin2016}
E.O. Kiktenko, A.S. Trushechkin, M.N. Anufriev, N.O. Pozhar, and A.K. Fedorov,
{\href{https://dx.doi.org/10.5281/zenodo.200365}{Post-processing procedure for quantum key distribution systems (2016)}}.

\bibitem{LDPC}
R. Gallager, 
{\href{https://coding.yonsei.ac.kr/gallager-ldpc.pdf}{IRE Trans. Inf. Theory {\bf 8}, 21 (1962)}}.

\bibitem{LDPC2}
D.J.C. MacKay, 
{\href{https://dx.doi.org/10.1109/18.748992}{IEEE Trans. Inf. Theory {\bf 45}, 399 (1999)}}.

\bibitem{MartnezMateo2010}
J. Mart{\'{\i}}nez{-}Mateo, D. Elkouss, and V. Martin,
{\href{https://dx.doi.org/10.1109/LCOMM.2010.101810.101384}{IEEE Comm. Lett. {\bf 14}, 1155 (2010)}}.

\bibitem{Elkouss2009}
D. Elkouss, A. Leverrier, R. Alleaume, and J.J. Boutros
{\href{https://dx.doi.org/10.1109/ISIT.2009.5205475}
{in {\it Proceedings of the IEEE International Symposium on Information Theory}, Seoul, South Korea, (2009), p. 1879}}.

\bibitem{Elkouss2012}
D. Elkouss, J. Mart{\'{\i}}nez{-}Mateo, and V. Martin,
{\href{https://dx.doi.org/10.1109/WCL.2012.082712.120531}{IEEE Wireless Comm. Lett. {\bf 1}, 585 (2012)}}.

\bibitem{Krovetz}
T. Krovetz and P. Rogaway, 
{\href{https://dx.doi.org/10.1007/3-540-45247-8_7}{Lect. Notes Comp. Sci. {\bf 2015}, 73 (2001)}}.

\bibitem{Gisin3}
N. Walenta, A. Burg, D. Caselunghe, J. Constantin, N. Gisin, O. Guinnard, R. Houlmann, P. Junod, B. Korzh, N. Kulesza, M. Legr\'e, 
C.C.W. Lim, T. Lunghi, L. Monat, C. Portmann, M. Soucarros, P. Trinkler, G. Trolliet, F. Vannel, and H. Zbinden,
{\href{https://dx.doi.org/10.1088/1367-2630/16/1/013047}{New J. Phys. {\bf 16} 013047 (2014)}}.

\end{thebibliography}
\end{document}